\documentclass[aps,prd,twocolumn,groupedaddress,nofootinbib]{revtex4}

\usepackage{amssymb}
\usepackage{epsfig}

\begin{document}

%\preprint{}

\title{Varying speed of light cosmology from a stringy short distance cutoff}


\author{Jens C. Niemeyer}
 \email[]{jcn@mpa-garching.mpg.de}
%\thanks{}
\affiliation{Max-Planck-Institut f\"ur Astrophysik,
Karl-Schwarzschild-Str.~1, D-85748 Garching, Germany}


%\date{\today}

\begin{abstract}
It is shown that varying speed of light cosmology follows
from a string-inspired minimal length uncertainty relation. 
Due to the reduction of the available phase space volume
per quantum mode at short wavelengths, the equation of state of
ultrarelativistic particles stiffens at very high densities. This
causes a stronger than usual deceleration of the scale factor
which competes with a higher than usual propagation speed of the
particles. Various
measures for the effective propagation speed are analyzed: the
group and phase velocity in the high energy tail, the thermal average
of the group and phase
velocity, and the speed of sound. Of these three groups, only the
first provides a possible solution to the cosmological horizon problem.
\end{abstract}

%\keywords{}

\maketitle

\section{\label{intro} Introduction}

Varying speed of light (VSL) cosmology  was conceived as an
alternative solution to the cosmological horizon problem
\cite{M93a,M93b,AM99,M00}. In 
contrast with inflation, where the comoving horizon scale shrinks as a
consequence of the accelerating growth of the scale factor, VSL keeps
the scale factor dynamics of the standard hot big bang unchanged but
assumes that the speed of light, $c(t)$, is a decreasing function of 
of cosmic time $t$. The horizon problem can be solved provided that
$c(t)$ drops sufficiently steeply for some time. In the original
formulation, $c(t)$ was treated as a free parameter of the theory.

In Ref.~\cite{AM01}, it was noted that VSL can effectively be obtained
from the thermodynamical properties of theories with nonlinear
dispersion relations which, in turn, were motivated in \cite{AM01} by
noncommutative geometry. While there are indications from string/M-theory
that noncommutativity of spatial coordinate operators may be relevant
near the string scale, we are lacking a simple, concrete realization
of noncommutative geometry that can be used in the cosmological
context. Hence one is forced to reduce its effects to a modified
dispersion relation.

A very similar yet more specific approach is taken in this work. The
starting point is an observation that was made in numerous studies of
quantum gravity and string theory: under rather general assumptions,
one can show that there exists a minimal length scale that can be
probed by experiments whereas the resolution of shorter distances is
prohibited by quantum gravitational effects
(e.g., \cite{GM88,ACV89,G95,W96}). Examined from the 
point of view of the low energy effective theory, it appears that
quantum gravity modifies the Heisenberg uncertainty relation to
\begin{equation}
\label{heisenberg}
\Delta x \Delta p \ge \frac{\hbar}{2} \left(1 + \beta (\Delta p)^2 +
\dots \right)
\end{equation}
which gives rise to a minimal short distance uncertainty $\Delta x_{\rm min}
= \hbar \sqrt{\beta}$ even for infinite $\Delta p$. 
For practical purposes, $\Delta x_{\rm min}$ can be identified
with either the string scale or the Planck length, where the former
could be a few orders of magnitudes larger than the latter. 
The correction
may of course originate from complicated dynamics of the underlying
fundamental theory. On the other hand, it can be modeled by equivalent
corrections to the canonical commutation relation:
\begin{equation}
\label{commutation}
[{\bf x},{\bf p}]= i \hbar (1 + \beta {\bf p}^2 + \dots)
\end{equation}
as discussed in \cite{K94}. Indeed, the uncertainty relation above
belongs to class of only very few types of short-distance structures
of space-time that are admitted under very general assumptions
\cite{K98}. Hilbert space representations of Eq.~\ref{commutation} were
employed, for instance, for regularizing field theory \cite{KM97} and, more
recently, for analyzing the impact of short distance uncertainty on
the predictions of inflation \cite{K00,KN01}.

The finite spatial localization of particles also changes their
thermodynamic behavior at very high density. This has already been
observed in \cite{L00,K01} where some cosmological consequences have been
pointed out. However, a complete analysis of this theory in the
context of VSL cosmology has not been done and shall be provided in
this work. The most relevant features of the minimal length uncertainty
theory are summarized in Sec.~\ref{Ktheory}, and are used to derive the
expressions for the energy density and pressure of a radiation fluid
in Sec.~\ref{eos}. In Sec.~\ref{vsl}, the cosmological implications
are discussed and a further outlook is presented in Sec.~\ref{concl}.

\section{\label{Ktheory} Quantum theory with minimal short distance
uncertainty}  

An explicit example for the choice of commutation relations in three
spatial dimensions that maintains translation and rotation invariance
is given in \cite{KM97}:
\begin{eqnarray}
[{\bf x}_i, {\bf p}_j] & = & i \hbar \left(\frac{\beta {\bf p}^2}{(1+2
\beta {\bf p}^2)^{1/2} - 1}\delta_{ij} + \beta {\bf p}_i{\bf
p}_j\right) \\
~[{\bf x}_i, {\bf x}_j] & = & 0 \\
~[{\bf p}_i, {\bf p}_j] & = & 0
\end{eqnarray}
Demanding translation and rotation invariance, this choice is unique
to first order in $\beta$ \footnote{Note that in \cite{L00}, a different choice for
the commutation relations was made which is noncommutative in the
spatial coordinates and breaks translation invariance.}. It does,
however, violate Lorentz invariance 
and thereby specify a preferred frame which, in the present context,
coincides with the cosmological rest frame. 

Owing to the finite short distance uncertainty, the usual Hilbert
space representation in terms of positions eigenstates is no longer
available (in contrast with the momentum representation $|p\rangle$, which still
is). An alternative representation is found by introducing the
translators ${\bf T}_i$ with eigenstates $|\kappa \rangle$ obeying ${\bf
T}_i |\kappa \rangle = \kappa_i/i \hbar |\kappa \rangle$, $i=1\dots3$,
and $\kappa^2 < 2/\beta$ \cite{KM97}. The projection onto momentum space is given by
\begin{equation}
\label{trafo}
\langle \kappa | p \rangle = \delta\left(p_i - \frac{\kappa_i}{1 - \beta
\kappa^2 / 2}\right)\,\,.
\end{equation}
In terms of $\psi(\kappa) = \langle \kappa | \psi \rangle$, the operator
representations and the scalar product become:
\begin{eqnarray}
{\bf x}_i \psi(\kappa) & = & i \hbar \partial_{\kappa_i} \psi(\kappa)\\
{\bf p}_i \psi(\kappa) & = & \frac{\kappa_i}{1-\beta \kappa^2/2} \psi(\kappa)\\
\langle \psi_1 | \psi_2 \rangle & = & \int\limits_{\kappa^2 < 2/\beta} {\rm d}^3 \kappa\,\,
\psi_1^\ast(\kappa) \psi_2(\kappa)\,\,.
\end{eqnarray}
The same representation was employed in \cite{K00,KN01}. 

As demonstrated in \cite{KM97}, it is now possible to construct the
closest analogue to the position representation by replacing the usual
position eigenstates with states of maximal localization around a
given position, $| x^{\rm ml} \rangle$. Their $\kappa$-space
representation can be constructed by applying the 
translation operator $e^{x{\bf T}}$ to the maximally localized field
around the origin, and therefore it varies as $\sim e^{-ix\kappa/\hbar}$. Using
(\ref{trafo}), one finds the quasi-position representation of the
plane wave with momentum $p$:
\begin{equation}
\langle p | x^{\rm ml} \rangle = N(p^2)\,\, \exp\left(\frac{-ix}{\hbar}\frac{p
\sqrt{1+2\beta p^2}-1}{\beta p^2}\right)\,\,,
\end{equation}
where $N(p^2)$ is a normalization coefficient.
It is now clear that $\kappa/\hbar$ acts as a substitute of the usual
wavenumber $k$, with the crucial difference that it only allows a
minimum wavelength $\lambda_{\rm min} = \pi \hbar (2 \beta)^{1/2}$ for
$p \to \infty$, as required at the onset. This will become important
for the computation of the thermodynamic variables in the next
section. 

\section{\label{eos} Equation of state of a radiation fluid}

In order to derive the energy density and pressure of an ideal gas of
photons (with straightforward generalization to other
ultrarelativistic particles), the grand canonical formalism can be
used (e.g. \cite{R80}). The grand potential is defined as
\begin{equation}
\Omega = -k_{\rm B}T \,\ln{\cal Z}\,\,,
\end{equation}
where ${\cal Z}$ is the grand partition function,
\begin{equation}
{\cal Z} = {\rm Tr}\, e^{-({\bf H} - \mu {\bf N})/k_{\rm B}T}\,\,,
\end{equation}
${\bf H}$ is the Hamiltonian, ${\bf N}$ is the number operator, $k_{\rm B}$
is Boltzmann's constant, and $T$ is the temperature. Evaluating the
trace in the number representation for bosons, one finds as usual
\begin{equation}
\label{om1}
\Omega =  k_{\rm B}T \, \sum\limits_l \,\ln\left(1 -
e^{-(E^l-\mu)/k_{\rm B}T}\right)\,\,.
\end{equation}
The summation is over all allowed momentum states $p^l$ and $E^l = c
|p^l|$ is the energy of a relativistic particle in this state. 

At this point, one usually transforms the sum into an integral over
the three-dimensional wavenumber $k$ by partitioning
space into boxes of volume $V=L^3$ and demanding $L$-periodicity of
plane waves with momentum $p = \hbar k$ in the position
representation. In the present framework, a position representation is
unavailable. Instead, the continuum limit is now most conveniently taken in
$\kappa$-space: requiring the periodicity of $e^{ix\kappa/\hbar}$ yields
$\kappa = 2 \pi \hbar l/L$, where $l \in M$ labels the discrete quantum state
and $M = \{l \in \mathbb{Z}^n \,|\, l^2 < (L/\lambda_{\rm
min})^2\}$. For sufficiently large $V$, the energy levels are nearly
continuous and Eq.~(\ref{om1}) can be written as 
\begin{equation}
\label{om2}
\Omega = \frac{g\, V \,k_{\rm B}T}{(2 \pi \hbar)^3} \int\limits_{\kappa^2 < 2/\beta}
{\rm d}^3 \kappa \,\ln\left(1 - e^{-[E(\kappa)-\mu]/k_{\rm B}T}\right)
\end{equation} 
with $E(\kappa) = c |p(\kappa)|$ and $g=2$ for photons. After transforming
the integral to energy 
space and writing it in terms of the dimensionless variables $\epsilon
= E/E_\beta$, $\tilde \mu = \mu/E_\beta$, and $\tau = k_{\rm
B}T/E_\beta$, where $E_\beta = c 
\beta^{-1/2}$ is the characteristic energy scale where quantum
gravitational effects become relevant, one obtains
\begin{equation}
\Omega = \frac{4 \pi\, g\, V\, E_\beta^4\, \tau}{(2 \pi \hbar\,
c)^3}\,\int\limits_0^\infty {\rm d}\epsilon \, J(\epsilon) \,
\epsilon^2 \, \ln\left(1 - e^{-[\epsilon-\tilde\mu]/\tau}\right)\,\,.
\end{equation}
Here, $J$ is the Jacobian determinant of the transformation $\kappa_i
\to p_i = \kappa_i (1 - \beta \kappa^2/2)^{-1}$, $p_i \in \mathbb{R}$,
given in terms of $\epsilon$ by
\begin{equation}
J(\epsilon) = \frac{2}{\epsilon^6}\left(2 + \epsilon^2 - \frac{2 +
3\epsilon^2}{\sqrt{1 + 2 \epsilon^2}}\right)\,\,.
\end{equation}
$J(\epsilon)$ essentially contains all of the information about the
modification of the phase space volume that results from the short
distance cutoff. 

One can now derive the thermodynamic variables from the grand
potential in the standard way. Using $\rho(\tau) = V^{-1}\,
\partial_{\tau^{-1}}\, \tau^{-1} \Omega$ for the energy density,
$P(\tau) = -\partial_V \,\Omega$ for the pressure, $N(\tau) = -V^{-1}\,
\partial_\mu\, \Omega$ for the number density and setting $\tilde \mu =
0$ for ultrarelativistic particles, one finds:
\begin{eqnarray}
\rho(\tau)& = & \frac{4 \pi\, g\, E_\beta^4}{(2 \pi \hbar\,
c)^3}\,\int\limits_0^\infty {\rm d}\epsilon \, J(\epsilon) \,
\frac{\epsilon^3}{e^{\epsilon/\tau}-1}\\
P(\tau)& = & -\frac{4 \pi\, g\, E_\beta^4\, \tau}{(2 \pi \hbar\,
c)^3}\,\int\limits_0^\infty {\rm d}\epsilon \, J(\epsilon) \,
\epsilon^2 \, \ln\left(1 - e^{-\epsilon/\tau}\right)\\
N(\tau)& = & \frac{4 \pi\, g\, E_\beta^3}{(2 \pi \hbar\,
c)^3}\,\int\limits_0^\infty {\rm d}\epsilon \, J(\epsilon) \,
\frac{\epsilon^2}{e^{\epsilon/\tau}-1}
\end{eqnarray}

It is useful to study the asymptotic behavior of the integrals for
very small and very large $\tau$ (e.g., \cite{H91}). In the case of
$\rho(\tau)$, rescaling the integrand by substituting
$u=\epsilon/\tau$ shows that the dominant contribution comes from
$u=O(1)$ for $\tau \to 0$. One can therefore write $J = 1 + O(u^2
\tau^2)$ and find the usual result $\rho(\tau) \sim \tau^4$ for $\tau
\to 0$, as expected. For large $\tau$, we first note that
$J(\epsilon)$ drops as $\epsilon^{-4}$ for large
$\epsilon$. Hence, for $\epsilon = O(1)$ the integrand is
$O(\tau)$ and so is the integral, whereas for $\epsilon = O(\tau)$,
the integrand is $O(\tau^{-1})$ and the integral is $O(1)$. The
dominant contribution therefore comes from $\epsilon = O(1)$ so that
the exponential can be approximated by $1 + \epsilon/\tau + O(\epsilon^2
\tau^{-2})$, yielding
\begin{equation}
\rho_{\tau \to \infty}(\tau) \simeq \frac{4 \pi\, g\, E_\beta^4}{(2 \pi \hbar\,
c)^3}\,\tau\,\int\limits_0^\infty {\rm d}\epsilon \, J(\epsilon)\,
\epsilon^2 \,\,,
\end{equation}
where the integral evaluates to $2^{3/2}/3$.

In the case of $P(\tau)$, partial integration yields
\begin{equation}
P(\tau) =  \frac{4 \pi\, g\, E_\beta^4}{3\,(2 \pi \hbar\,
c)^3}\,\int\limits_0^\infty {\rm d}\epsilon \, G(\epsilon) \,
\frac{\epsilon^3}{e^{\epsilon/\tau}-1}\,\,,
\end{equation}
where $G(\epsilon)$ is defined as
\begin{eqnarray}
G(\epsilon) & = & \frac{3}{\epsilon^3}\,\int {\rm d}\epsilon\,
J(\epsilon)\,\epsilon^2\nonumber\\
& = & \sqrt{1 + 2 \epsilon^2}\,J(\epsilon)\,\,.
\end{eqnarray}
For small $\tau$, $G(\epsilon) = 1 - O(\epsilon^2)$ behaves like
$J(\epsilon)$, and one obtains the expected result $P = \rho/3$. On
the other hand, $G \sim \epsilon^{-3}$ drops more slowly for $\tau \to
\infty$ than $J(\epsilon)$. It is only for this reason that the ratio
of pressure and energy density, and hence the dynamics of the
scale factor, differs from its usual behavior at high temperatures, as
will be demonstrated in Sec.~\ref{vsl}. Indeed, for $\epsilon =
O(\tau)$, the integrand is now $O(1)$ and the integral is $O(\tau)$,
contributing equally strongly to the integral as $\epsilon = O(1)$. In
this case, a good approximation is given by 
\begin{eqnarray}
P_{\tau \to \infty}(\tau) & \sim & \int \limits_{O(1)}^{O(\tau)} {\rm
d}\epsilon \, \frac{\tau}{\epsilon}\nonumber\\
& = & \tau \, (A + B \log \tau)\,\,.
\end{eqnarray}

The cosmological evolution of $\rho$ is determined, via the energy conservation
equation, by the equation of state parameter $w = P/\rho$. For small
$\tau$, one recovers the usual equation of state, $w = 1/3$, whereas
$\tau \to \infty$ gives $w_{\tau \to \infty}(\rho) = C + D \log
\rho_{\tau \to \infty}$. The 
same form of $w$ was discovered and analyzed in \cite{AM01}, hence one can
expect a similar result for the cosmological evolution. As shown in
Sec.~\ref{vsl}, this is indeed true. The result for $w(\rho)$ was
confirmed numerically by inverting $\rho(\tau)$ in order to evaluate
$P(\tau)$. A good high-density fit is provided by $C \simeq -1.2$ and
$D \simeq 2.3$.

Finally, using the same arguments as for $\rho(\tau)$, the expression
for $N(\tau)$ yields the standard result for $\tau \to 0$ while $\tau
\to \infty$ gives
\begin{equation}
N_{\tau \to \infty}(\tau) \simeq \frac{4 \pi\, g\, E_\beta^3}{(2 \pi \hbar\,
c)^3}\,\tau\,\int\limits_0^\infty {\rm d}\epsilon \,J(\epsilon)\, \epsilon =
\frac{2 \pi\, g\, E_\beta^3}{(2 \pi \hbar\,c)^3}\,\tau\,\,.
\end{equation}

All of the preliminaries are now in place to examine the cosmological
impact of the modified equation of state, which will proceed in close
analogy with Ref.~\cite{AM01}.

\section{\label{vsl} Cosmological implications}

\begin{figure}
\epsfxsize=0.5\textwidth
\epsfbox{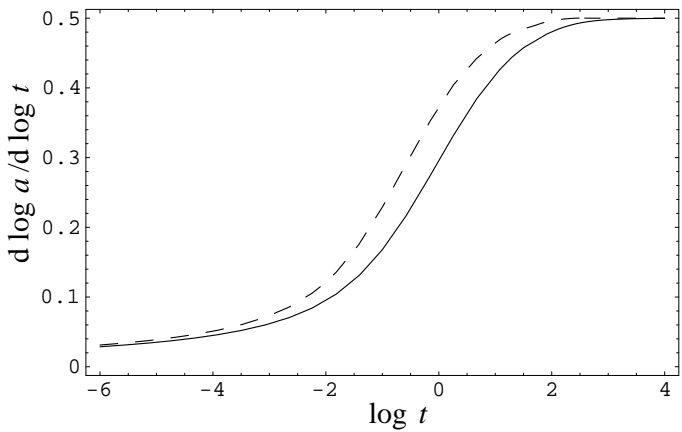}
\caption{\label{f1} Comparison of the effective time exponent of the
numerical solution (solid line) and the approximation $2/3(1+w(\rho))$
(dashed line) as a function of time, adjusted for a
shift of the time of the cosmic singularity.}
\end{figure}

The evolution of $\rho(t)$ and the scale factor $a(t)$ in the very
early universe are determined by the Friedmann and energy conservation
equations with negligible curvature and cosmological constant terms,
given by
\begin{eqnarray}
\label{fried}
\left(\frac{\dot a}{a}\right)^2 & = & \frac{\rho}{3 M_{\rm Pl}^2}\\
\label{cons}
\frac{\dot \rho}{\rho} & = & - \frac{3 \dot a}{a}\left(1+w(\rho)\right)\,\,,
\end{eqnarray}
where a dot corresponds to a derivative with respect to cosmic time
and $M_{\rm Pl} = (8\pi G)^{-1/2}$ is the reduced Planck
mass. These equations can only be solved numerically (starting with
the known low-temperature solution and integrating backward), using the
numerical evaluation of $w(\rho)$. However, if $w(\rho)$ varies
sufficiently slowly with $\rho$, one can approximate the solution for
the scale factor by the usual one for constant $w$, with $w$ replaced
by $w(\rho)$ and an offset for the origin of the time coordinate $t_0$:
\begin{equation}
\label{aprsol}
a(t) \simeq (t-t_0)^{2/3(1+w(\rho))}\,\,.
\end{equation}
Fig.~\ref{f1} demonstrates that this is a good approximation at all
times except for the brief period where $w$ transitions from $1/3$ to
the logarithmic high-temperature behavior.

Eq.~(\ref{aprsol}) captures the relevant scale
factor dynamics in our model: at very high density, the deceleration
$\sim \ddot a/a H^2$ is very large, giving rise to a
smaller than usual Hubble rate $H = \dot a/a$, whereas the standard
scale factor evolution is recovered at densities far below the string scale.

The high-density evolution of $\rho$ as a function of the scale factor
can be obtained directly by solving Eq.~(\ref{cons}) with $w(\rho)
\simeq w_{\tau \to \infty}$, yielding
\begin{equation}
\log \rho_{\tau \to \infty}(a) = E\,a^{-3 D} - \frac{C+1}{D}\,\,.
\end{equation}
This was also shown in \cite{AM01}.

The question whether the cosmological horizon problem can be solved in
this model hinges crucially on the choice for the effective
propagation speed of information, $c_{\rm eff}$. The VSL mechanism
works if the comoving horizon scale,
\begin{equation}
r_{\rm h} = \frac{c_{\rm eff}}{a H} =  \frac{c_{\rm eff}}{\dot a}\,\,,
\end{equation}
declines for some period of time before turning around to its current
growing behavior. 

A case can be made that the most
suitable measure for $c_{\rm eff}$ is given by the speed of sound,
$c_{\rm s}$, of the radiation 
fluid rather than by the group or phase velocity of individual
photons. It is given by
\begin{equation}
c_{\rm s}^2 = \left(\frac{\partial P}{\partial \rho}\right)_{S=const} = w + \rho
\frac{\partial w}{\partial \rho}\,\,.
\end{equation}
In the high-density limit, $c_{\rm s}$ scales like $(\log
\rho)^{1/2} \sim a^{-3/2}$, whereas $\dot a \sim a \rho^{1/2} \sim a
\exp(1/2a^3)$ which clearly wins as $a \to 0$. Consequently, $r_{\rm
h}$ is always a growing function of time in this case and the horizon
problem remains. 

The group and phase velocities are given by
\begin{eqnarray}
c_{\rm g}(\epsilon) =  \frac{\partial E}{\partial |\kappa|}& = &
\frac{c}{2}\left(1 + 2\epsilon^2 + \sqrt{1 + 2\epsilon^2}\right)\\
c_{\rm ph}(\epsilon) = \frac{E}{|\kappa|}& = & 
\frac{c}{2}\left(1  + \sqrt{1 + 2\epsilon^2}\right)\,\,.
\end{eqnarray}
As pointed out in \cite{AM01}, the evaluation
of $\epsilon$ in these expressions is somewhat ambiguous. The bulk of
all photons will 
populate a region around the peak in the modified Planck distribution,
which saturates at $\tau \sim 1$. After saturation, the group and phase
velocities of these photons will cease growing as a function of
increasing density. However, the authors of \cite{AM01} argue that the
``fast tail'' of the Planck distribution may provide the required
causal contact, and they propose to evaluate $c_{\rm g}$ and $c_{\rm
ph}$ at $\epsilon \simeq \tau \simeq \rho_{\tau \to \infty}$. In this
case, $c_{\rm g} \sim \rho^2$ and $c_{\rm ph} \sim \rho$ at high
densities, so that $r_{\rm h}$ indeed declines as a function of time
(consistent with \cite{AM01}). Consequently, this choice of effective
velocity solves the horizon problem, provided that the model assumptions
made in Sec.~\ref{Ktheory} remain valid well into the regime $\tau \gg
1$.  

As a final possibility, one can consider a thermal average of
$c_{\rm g}$ (or $c_{\rm ph}$), defined for instance as the particle number
weighed value:
\begin{eqnarray}
\langle c_{\rm g,ph} \rangle(\tau) & = & N^{-1}(\tau)\,\frac{4 \pi\, g\,
E_\beta^3}{(2 \pi \hbar\, 
c)^3}\,\int\limits_0^\infty {\rm d}\epsilon \,
\frac{ K_{\rm g,ph}(\epsilon) \,\epsilon^2}{e^{\epsilon/\tau}-1}\\
 K_{\rm g,ph}(\epsilon) & = & c_{\rm g,ph}\,J(\epsilon)\,\,.
\end{eqnarray}
The asymptotic properties of $N(\tau)$ were discussed in
Sec.~\ref{eos}. Noting that $K_{\rm g,ph} = 1 - O(\epsilon^2)$ for
small $\epsilon$ one finds $\langle c_{\rm g,ph} \rangle(\tau) \simeq
c$ as required. On the other hand,  $K_{\rm g} \sim \epsilon^{-2}$ for
large $\epsilon$, whereas $K_{\rm ph} \sim \epsilon^{-3}$. The
situation for the integral over  $K_{\rm g}$ at $\tau \to \infty$ is
therefore analogous to that of $P_{\tau \to \infty}$, yielding the
same logarithmic growth. The normalization by $N(\tau)$ gives rise to
the same behavior as found earlier for $w_{\tau \to \infty}$, i.e.
$\langle c_{\rm g} \rangle_{\tau \to \infty} \sim \log \rho$. In
contrast, the integral over  $K_{\rm ph}$ at $\tau \to \infty$ is
analogous to that of $\rho_{\tau \to \infty}$, hence the normalized
result for $\langle c_{\rm ph} \rangle_{\tau \to \infty}$ asymptotes
to a constant independent of $\tau$. Explicitly, one finds  $\langle
c_{\rm ph} \rangle_{\tau \to \infty} = 2c$.

To summarize, the thermally averaged group or phase velocities grow at
best logarithmically. This is insufficient to solve the horizon
problem by means of the VSL mechanism, as argued above for $c_{\rm
s}$. 

\begin{figure}
\epsfxsize=0.5\textwidth
\epsfbox{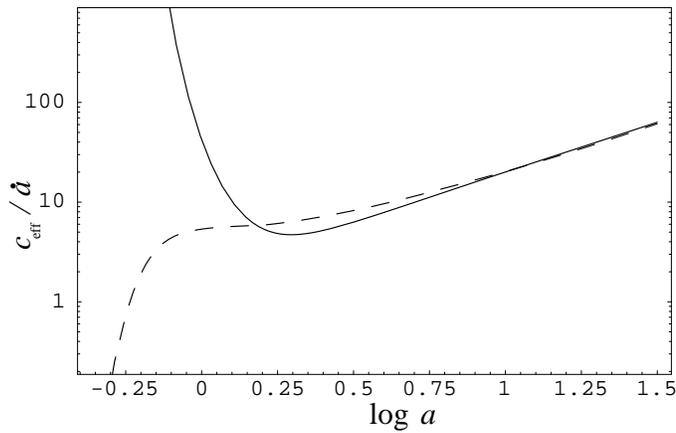}
\caption{\label{f2} Comoving horizon distance as a function of scale
factor for the ``fast tail'' group velocity $c_{\rm g}$ (solid line)
and the thermally averaged one, $\langle c_{\rm g} \rangle$ (dashed
line). While the former can solve the horizon problem, the latter
cannot. At late times, both converge onto the radiation dominated
solution $\sim t^{1/2}$.}
\end{figure}

Fig.~\ref{f2} shows the size of the comoving horizon as a
function of time, computed numerically using the ``fast tail'' group
velocity and the thermally averaged one.

\section{\label{concl} Summary and Outlook}

The analysis presented in this work is based on a rather simple
premise: The 
spatial separation that can be probed by high-energy particles has a
finite minimum value. Without referring to any specific quantum
gravitational effect in particular, it is a very general
model for the physics of relativistic particles at densities that
occurred in the very early universe. It is very interesting that such
a simple modification may provide an alternative solution to the
horizon problem without resorting to the inflationary paradigm. Of
course, one must now think of possibilities to resolve other quandaries
that inflation has proved to be so helpful with, such as the flatness
and relics problems and, perhaps most importantly, the generation of
scale invariant fluctuations. Some ideas in this direction have
already been put forward in the context of VSL \cite{AM99,AM01}, and
they can be readily generalized to the model in this work.

As always, many open questions remain. One of them, which may be
answered by a more detailed analysis of transport processes in the
early universe, is the correct choice of the effective communication
speed. On a more fundamental level, one may question the validity of
the minimal distance uncertainty principle as early as may be required
to solve the horizon problem. In other words, it is possible (and even
likely) that additional, quantum gravitational degrees of freedom will
be excited in the relevant regime of $\rho$ that will drastically
change the equation of state. Unfortunately, the answer to this
question might remain elusive until the fundamental theory is known in
full detail.

\begin{acknowledgments}
I am happy to thank Achim Kempf and Joao Magueijo for valuable comments.
\end{acknowledgments}

\bibliography{early_universe}

\end{document}